\documentclass[epj,referee]{svjour}
\usepackage{graphics}
\begin{document}
\title{Size-dependent effects in solutions of small metal nanoparticles}
\author{Vitaliy N. Pustovit\inst{1,2}, Tigran V. Shahbazyan\inst{2} \and Leonid G. Grechko 
\inst{1} % \thanks is optional - remove next line if not needed
}                     % Do not remove
\institute{Laboratory of Surface Physics, Institute of Surface
  Chemistry, Kyiv 03164, Ukraine \and Department of Physics, Jackson State University, Jackson, MS  39217, USA}
\date{Received: date / Revised version: date}
\abstract{
A new theoretical approach for the calculation of optical properties of complex
solutions is proposed. It is based on a dielectric matrix $\varepsilon_m$ with included small
metallic inclusions (less than 3 nm) of spherical
shape. We take into account the mutual interactions between the inclusions and the quantum finite-size effects. On the
basis of the effective medium model, TDLDA and Kohn-Sham theories, some analytical expressions for the effective dielectric permittivity of the solution are obtained.
 % end of PACS codes
} %end of abstract
\maketitle
\section{Introduction}
\label{intro}
The optical properties of the solutions with small metal particles have attracted much attention in recent years. A number of theories have been proposed for describing the anomalous far-infrared  absorption enhancement in this kind of systems \cite{shalaev}. Most authors agree that a direct account of the interaction between particles in such systems has a significant influence on the behavior of the absorption spectra of these systems \cite{claro}. In many previous works mutual interactions between particles were taken into account \cite{claro}-\cite{felder-phys}, though until now the influence of quantum-size effects for interacting small metal nanoparticlers on the absorption spectra remains uninvestigated.
In this paper, we show influence of the quantum-size effects on the optical properties of
the solutions (matrix disperse systems) based on dielectric matrix and embedded  in  small metallic inclusions (less 3 nm) of spherical shape.
We demonstrate that level of mutual interactions between metallic inclusions 
is increased, especially for the small size nanoparticles where quantum-size effects 
became most pronounced \cite{pust-prb}.
The underlying mechanism of enhancement is related to the difference in the density profiles of 
{\em sp}-band and {\em d}-band electrons near the nanoparticle
boundary. Specifically, the {\em localized} {\em d}-electrons are
mainly confined within nanoparticle classical volume while the
wave-functions of {\em delocalized} {\em sp}-electrons extend outside
of it. This spillover leads to a larger {\em effective} radius for
{\em sp}-electrons \cite{persson-prb85} and, thus, to the existence of
a surface layer with diminished {\em d}-electron population.
\cite{liebsch-prb93,liebsch-prb95} As a result, in the
surface layer, the screening of {\em sp}-electrons by {\em d}-band
electron background is reduced, leading to the enhancement of the local field 
and correspondingly to the mutual interactions between neighboring metallic particle.

The paper is organized as follows. In Section \ref{Maxwell} we
 develop model of extended Maxwell-Garnett approximation \cite{maxwell} which will includes 
all mutual multipolar interactions between neighbor metal inclusions \cite{pust-apl}. 
In Section \ref{results} we calculate absorption spectra for the solution of Au nanoparticles.
Discussion of our numerical results is presented in the same Section. 
In Appendix we propose original way for derivation of quantum multipolar polarizability of metal nanoparticle.
Section \ref{sec:conc}
concludes the paper.

\section{Extended Maxwell-Garnett approximation}
\label{Maxwell}
Consider a system of N non-overlapping spherical metallic particles randomly distributed 
in the background media and interacting with each other and with an incident plane monochromatic wave, 
whose wavelength $\lambda$   is assumed to be much larger than the average size of the particles in the cluster
 (i.e., electrostatic approximation) \cite{bohren}. The particles have approximately the same radius $R$ and embedded in a dielectric continuous matrix with dielectric constant $\varepsilon_m$ . 
Due to the influence of the external field as well as the interaction between particles in the solution, all particles in the system
 will obtain some polarization. The total effective dipole moment of particle $i$, accounting mutual neighboring many-particle interactions, can be expressed as
\begin{eqnarray}
\label{polarisation}
{\bf p}_{i}(1,2...N)={\bf p}^{(0)}_{i} + \sum^{N}_{j \neq i} {\bf p} _{ij} + ...,
\end{eqnarray}
where is a dipole moment of an isolated particle  $i$ induced by the external field  ${\bf E_{0}}$. 
 The second term in the expansion (1) is defined by
\begin{eqnarray}
\label{polarisation2}
{\bf p} _{ij}={\bf p} (i,j) - {\bf p}^{(0)}_{i},
\end{eqnarray}
where ${\bf p} (i,j)$  is the dipole moment induced in particle  $i$ in the presence of a second particle $j$ . The second term in expansion (1-2) defines pair interactions between particles in the given mixture, the third three- particle interactions, and so on. We have restricted ourselves to considering the case of pair interactions between particles, because the inclusion of the third term will render the system mathematically hard to treat. 
As it is follow from the Maxwell equations, in the electrostatic approximation the average  Lorentz field is connected with average polarization in the system through the Green function equation
\begin{eqnarray}
\label{Green}
\langle{\bf F}(r)\rangle = {\bf E}_0 + \int G({\bf r} - {\bf r'}) \langle{\bf P}(r')\rangle dr'
\end{eqnarray}
From the other side, the averaged macroscopic polarization is linked with averaged Lorentz field in the solution as 
\begin{eqnarray}
\label{average_polar2}
\langle{\bf P}\rangle= \frac{3}{4 \pi} \frac{\hat\varepsilon-\varepsilon_m}{\hat\varepsilon+2\varepsilon_m} \langle{\bf F}\rangle 
\end{eqnarray}
where $\hat\varepsilon$ is a effective dielectric permittivity of the solution. 
Since we have restricted ourselves with first two terms of expansion (1),  the averaged macroscopic polarization of the solution becomes
\begin{eqnarray}
\label{average_polar}
\langle{\bf P}(1,2...N\rangle)={\bf p}^{(0)}_{i} +\frac{1}{V}  \sum_{i \neq j} {\bf p}_{i,j} ({\bf r}_i, {\bf r}_j) \Phi  ({\bf r}_i, {\bf r}_j)dr_j
\end{eqnarray}
where $V$ is a total system volume and $\Phi  ({\bf r}_i, {\bf r}_j)$ is a pair correlation function of the particle distribution in the solution.

In order to find expression for the pair dipole polarization ${\bf p}_{i,j}$  lets us solve two particle problem in the electrostatic approximation.

\begin{figure}
\resizebox{0.75\columnwidth}{!}{%
  \includegraphics{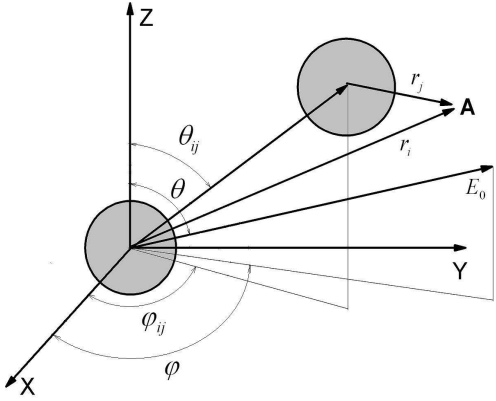}
}
%\vspace{10cm}       % Give the correct figure height in cm
\caption{Two metallic nanoparticles in the external field ${\bf E}_{0}$,}
\label{fig:size}       % Give a unique label
\end{figure}
First, we denote by $ {\bf r}_{i}$ the center of the sphere  $i$  and by $r$  an arbitrary point in the medium. 
The potential inside sphere $i$ can be expressed in the form:
\begin{eqnarray}
\label{potential_in}
\Psi_{in}^{(i)}=-{\bf E_{0}} \sum_{lm} C_{lm}^{(i)} |r-r_i|^l Y_{lm} (\hat {r-r_i}),
\end{eqnarray}
where  $\hat {r-r_i}$ is the unit vector along the direction $r-r_i$ and $Y_{lm} (\hat {r-r_i})$ 
are the scalar spherical harmonics (SSH) so $l$ is the polar order. The potential outside sphere  $i$ can be expressed in the form:
\begin{eqnarray}
\label{potential_out}
\Psi_{out}^{(i)}=-{\bf E_{0}} \sum_{lm} d_{lm} |r-r_i|^l Y_{lm} (\hat {r-r_i})
\nonumber\\
 - {\bf E_{0}} \sum_{lm} B_{lm}^{(i)} |r-r_i|^{-l-1} Y_{lm} (\hat {r-r_i})
\nonumber\\
- {\bf E_{0}}\sum_{j \neq i} \sum_{lm} B_{lm}^{(j)} |r-r_j|^{-l-1} Y_{lm} (\hat {r-r_j}),
\end{eqnarray}
where the first term is the potential of the external field, the second term is the potential created by the particle  $i$ at the point ${\bf r}$ , and the third term is the potential created by the rest of the particles of the solution at this point. Breaking up the potential inside the sphere into components in the two regions and applying standard boundary conditions at the three boundaries, plus use of
\begin{eqnarray}
\label{boundary}
d_{lm}=-\delta_{l1}\sqrt{\frac{2 \pi}{3}}\lbrace \sqrt{2} \delta_{m0} {\bf u}_z +i [\delta_{m1} + \delta_{m-1}] {\bf u}_y
\nonumber\\
+ [\delta_{m-1} - \delta_{m1}] {\bf u}_x\rbrace
\end{eqnarray}
where ${\bf u} = {\bf E_0}/E_0$, leads to the coupled set of equations
\begin{eqnarray}
\label{coupled}
\sum_{l'=1}^{\infty} T_{ll'}^{(m)} X_{l'm}= \delta_{l1};~~ l=1,2...;~~ m=-l,...l,
\end{eqnarray}
where
\begin{eqnarray}
\label{coupled2}
T_{ll'}^{(m)}= \frac{R^{2l+1}}{\alpha_l} \delta_{ll'} - (-1)^m(R/r_{ij})^{l+l'+1}
\left(
\begin{array}{ccc}
l+l'\\
l+m
\end{array}
\right)
\end{eqnarray}
where $R$ is a radius of nanoparticle, and $X_{lm}=B_{lm}R^{-(l+2)}$ with symbol
\begin{eqnarray}
\label{symbol}
\left(
\begin{array}{ccc}
l\\
m
\end{array}
\right) = \frac{l!}{m!(l-m)!},
\end{eqnarray}
Following the classical approach for polarizability of the single spherical particle we should apply an 
expression  which can be easily derived from the Mie theory \cite{bohren}
$ \alpha_l= R^{2l+1} l(\varepsilon-\varepsilon_m)/(l\varepsilon+(l+1)\varepsilon_m)$.
Although, if we consider small size nanoparticles (less $2nm$), that might be  necessary to take into 
account finite-size effects \cite{pust-prb}. The role of interband screening and $sp$ band spillover 
on the boundary of small particle will make a significant influence on the general multipolar 
polarizability. The quantum multipolar polarizabily in Eq.(\ref{coupled2})  can be defined 
in terms of induced density of electrons \cite{pust-prb}  $\delta n=\delta n_s + \delta n_d + \delta n_m$ 
due to influence of interband screening  and $sp$ band spillover as 
$\alpha_l^{q}= \alpha_l^{s}+\alpha_l^{d}$ where
\begin{eqnarray}
\label{alpha}
\alpha_l^{q}=  A_l R^{2l+1}=
\int_0^{\infty}dr'r'^{l+2}\delta n^{(l)}(r'),
\nonumber\\
\alpha_{l}^{d}
=
\frac{l R^{2l+1}}{\eta}\Bigl[ \lambda_d a^{2l+1} (1-\lambda_m) - \lambda_m 
\Bigr],
\nonumber\\
\alpha_{l}^{s}
=\int_0^{\infty}dr'r'^{l+2}\delta n_s^{(l)}(r')~~~~~~~~
\nonumber\\
-\frac{l(2l+1) R_d^{l+1} \lambda_d}{4 \pi \eta} \int_0^{\infty}dr'r'^2\delta n_s^{(l)}(r')
\Biggl[B_{l}(R_d,r')
\nonumber\\
-a^{l}(l+1)
\lambda_m B_{l}(R,r')\Biggr]\nonumber\\ + 
\frac{l(2l+1) R^{l+1} \lambda_m}{4 \pi \eta} \int_0^{\infty}dr'r'^2\delta n_s^{(l)}(r')
\Biggl[B_{l}(R,r')
\nonumber\\
-la^{l+1}\lambda_d B_{l}(R_d,r') \Biggr],~~~~~~~~~~~~~~
\end{eqnarray}
with coefficients given in Appendix A

\begin{figure}
\resizebox{0.75\columnwidth}{!}{%
  \includegraphics{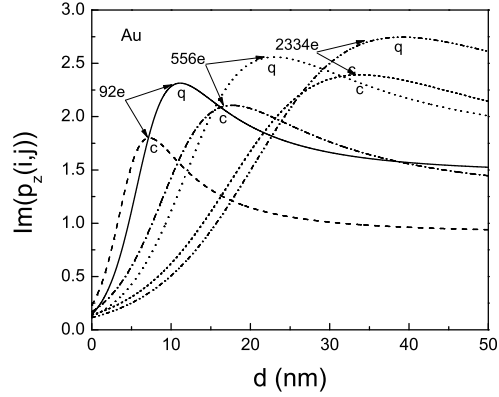}
}
%\vspace{5cm}       % Give the correct figure height in cm
\caption{Image part of the pair dipole interaction given in Eq.(\ref{pair2}) in $Z$ direction vs distance between two particles. 
The letters "c" and "q" defines application of the classical and TDLDA methods correspondingly for particles with 92e, $R=0.7nm$; 556e, $R=1.3nm$ and 2334e, $R=2nm$. }
\label{fig:distance}       % Give a unique label
\end{figure}

Finally, from the boundary equations plus  Eqs. (5-9) we obtain the expression for pair dipole moment
\begin{eqnarray}
\label{pair}
{\bf p}(i,j)= -{\bf E}_0 \sum_{m=1}^{1} B_{1m}^{(i)} Y_{1m}( {\bf r}_i-  {\bf r}_j)
\end{eqnarray}
We can now derive an expression for ${\bf p}_{i,j}$ as
\begin{eqnarray}
\label{pair2}
{\bf p}_{i,j}= R^3 [X_{10} n_z u_z + X_{11}(n_x u_x+ n_y u_y)] {\bf E}_0,
\end{eqnarray}
where ${\bf n}= {\bf r_{ij} }/r_{ij}$  and $r_{ij}=|r_i -r_j|$. 
Here coefficients $X_{10}$ and $X_{11}$ are obtained from the coupled system of equations Eq.(\ref{coupled}).
By selecting position of two particles along the axis $Z$ and parallel to external field $\bf E_0$, we can investigate dependence of the pair dipole 
interaction magnitude on the distance between two particles (see Fig ~\ref{fig:distance}).  Starting with small size of particle $92e$ or $R=0.7nm$ the maximum contribution to the absorption spectra will bring the particles located on relatively far distances from each other, i.e. $10$, $20$ and $40$ nm.  That conclusion lets us to assume that further in our calculations we can restrict ourselves only with the dipole $l=1$ interactions between particles, neglecting with higher orders
 multipoles, which are important for the close distances. 
After performing integration over distance between each pair of particle in  Eq.(\ref{average_polar}) with pair correlation function $\Phi(r_{ij})$ 
 and taking into account expression for the Lorentz field in Eq.(\ref{average_polar2}), we get a 
final expression for calculation of the effective dielectric permittivity of the solution with account of pair multipole interaction between metallic 
nanoparticles \cite{pust-apl} - \cite{pust-spie}.
\begin{eqnarray}
\label{main_form}
\frac{\hat\varepsilon+2\varepsilon_m}{\hat\varepsilon-\varepsilon_m}=\frac{1}{fA_1} ~~~~~~~~~~~~~~~~~~~~~~~~~~~~~~~~~~~~~~~
\nonumber\\
- \frac{1}{A_{1}^2 R^6} \int_{0}^{\infty}
r_{ij}^2 \Phi(r_{ij}) [\beta^{\parallel}(r_{ij}) + 2  \beta^ {\perp }(r_{ij})] dr_{ij},
\end{eqnarray}
where parameters
\begin{eqnarray}
\label{beta}
\beta^{\parallel}=X_{11}-A_1,~~~ \beta^ {\perp }=X_{10}-A_1,
\end{eqnarray}
where $A_1$ is given by the $l=1$ term in  Eq.(\ref{alpha}) and $f=\frac{4\pi}{3} R^3 n$ is a volume fraction of metallic particles in solution with concentration $n$.
The second term on the right side of Eq.(\ref{main_form}) is a contribution of the pair interaction between particles. Omitting this term, Eq.(\ref{main_form}) is easily reduced to the simple traditional form of the Maxwell-Garnett approximation \cite{maxwell}.
%%%%%%%%%%%%%%%%%%%%%%%%%%%%%%%%%%%%%%%%%%%%%%%%%%%%%%%%%%%%
\section{Numerical results and discussion}
\label{sec:num}
Below we present results of our calculations for Au nanoparticles with the same radius
ranging from 0.7 to 2 nm in a medium with dielectric constant $\varepsilon_m=1.77$. The surface plasmon (SP) resonance is positioned at $2.35$ $ Ev$,
far away from the interband transitions for gold \cite{palik-book}. Further we will consider absorption properties of the given solution with metallic inclusions in the SP frequency region.  The volume fraction  of nanoparticle inclusions in the solution is $f=0.1$ and we used the simplest approximation for the two particle distribution function
\begin{eqnarray}
\label{distribution}
\Phi(r_{ij})= \left\{
\begin{array}{ccc}
1 ~~~r_{ij}>3R\\
0 ~~~r_{ij}<3R
\end{array}
\right.,
 \end{eqnarray}
where minimal possible distance between two particles is selected to be at least more than one radius or $3R$ between particle centers.

\begin{figure}
\resizebox{0.75\columnwidth}{!}{%
  \includegraphics{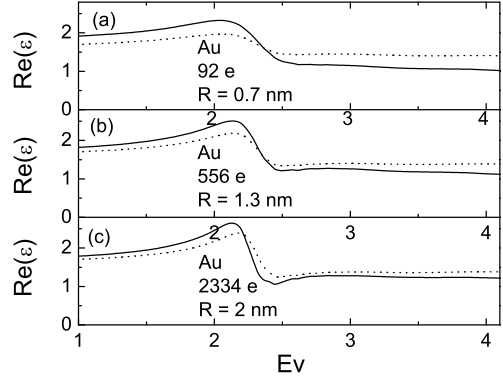}
}
%\vspace{5cm}       % Give the correct figure height in cm
\caption{Calculated real part of the medium effective dielectric constant in dipole approximation for solution with Au nanoparticles with 92,  556 and 2334 electrons 
at volume fraction $f=0.1$. The solid line corresponds to the calculations based on TDLDA method, and the dotted corresponds to classical approach.}
\label{fig:compare1}       % Give a unique label
\end{figure}
%

\begin{figure}
\resizebox{0.75\columnwidth}{!}{%
  \includegraphics{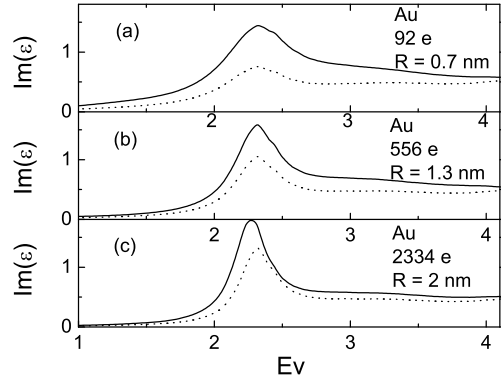}
}
%\vspace{5cm}       % Give the correct figure height in cm
\caption{Calculated image part of the medium effective dielectric constant in dipole approximation for solution with Au nanoparticles with 92,  556 and 2334 electrons 
at volume fraction $f=0.1$. The solid line corresponds to the calculations based on TDLDA method, and the dotted corresponds to classical approach.}
\label{fig:compare2}       % Give a unique label
\end{figure}

To ensure spherical symmetry, only closed-shell "magic numbers" $N$ of electrons  92, 556 and 2334 were used \cite{ekardt-prb85}.
 For such sizes, the Au band-structure remains intact. The ground-state energy spectrum and wave functions were
obtained by solving the Kohn-Sham equations for the jellium model \cite{liebsch-prb93} with the Gunnarsson-Lundqvist exchange-correlation
potential \cite{lundqvist-prb77}. These results were used as input in the numerical solution of TDLDA system A(13). The effective radius of nanoparticles was taken in form $R=r_{s}N^{1/3}$, where parameter $r_{s}$, defined as radius in atomic units, is equal to 3 a.u. for Au.   
On Fig ~\ref{fig:compare1} and Fig ~\ref{fig:compare2} are shown calculated real and image parts of effective dielectric permittivity  of given solution obtained from Eq.\ (\ref{main_form}). Here we can compare absorption ($Im(\hat\varepsilon)$) computed with use of TDLDA quantum polarizability $\alpha_l^q$ (layer with sp-free electrons is included with parameter $\Delta=1$, see derivations in\cite{pust-prb}) with
absorption computed classically with polarizability $\alpha_l$.
 For all three cases (a)-(c) on Fig ~\ref{fig:compare1} and Fig ~\ref{fig:compare2}  results based on TDLDA  display higher level of absorption in the system, especially for the solution with small particles (like 92 e). In such case of small particle size the difference between two magnitudes can reach almost 2 times and significantly reduces
with increase of particle radius. This fact allows us to predict that in the solutions with small particles the level of absorption can be higher than it was expected before
based on the classical calculations. 
\section{Conclusions}
\label{sec:conc}
We proposed a theoretical method for calculation of optical properties of the complex solutions based on dielectric matrix $\varepsilon_m$
with included in small metallic inclusions of spherical shape with account of the dipole interaction between inclusions and quantum finite-size effects. 
We shown that optical properties of given solutions are very much dependent on the proper account of microscopic effects, in particular, electron screening
in the surface layer of nanoparticles. The exit of delocalized s- electrons out of the classical boundary leads to the additional local field enhancement and as a consequence to enhancement of the surface plasmon of nanoparticle. That correspondingly makes influence on the character of the dipole interaction between nanoparticles in solution and on the optical properties of this solution.

%This work was supported by ....

\appendix
\section{Derivation of quantum multipolar polarizability}
\label{results}

The full self-consistent potential 
$\Psi({\bf r})=\psi ({\bf r})+\psi_0({\bf r})$ satisfies Poisson equation

\begin{equation}
\label{poisson}
\Psi({\bf r})=\psi_0({\bf r}) + e^2 \int d^3r'
\frac{\delta N({\bf r}')}{|{\bf r}-{\bf r}'|},
\end{equation}
where $\psi_0({\bf r})=-e {\bf E}_0  \dot  {\bf r} $ is potential of the external electric field,
the induced density is comprised of {\em sp}-band, {\em d}-band
and medium contributions, 
$\delta N({\bf r})=
\delta N_s({\bf r})+\delta N_d({\bf r})+\delta N_m({\bf r})$.
The induced density
of {\em sp}-band electrons is determined from TDLDA equation presented below  and
contributions from d-electrons $\delta N_d({\bf r})$ and medium 
$\delta N_m({\bf r})$
 are back to $\Psi({\bf r})$ as 
\begin{eqnarray}
\label{back}
\delta N_d({\bf r})= \nabla
\bigl[\chi_d(r)\nabla \Psi({\bf r})\bigr],
\nonumber\\
\delta N_m({\bf r})= 
\nabla \bigl[\chi_m(r)\nabla \Psi({\bf r})\bigr],
 \end{eqnarray}
where
$\chi_d(r)=\frac{\epsilon_d-1}{4\pi} \,\theta(R_d-r)$ is the interband
susceptibility with the step function enforcing the boundary
conditions and, correspondingly, $\chi_m(r)=\frac{\epsilon_m-1}{4\pi}
\,\theta(r-R)$ is the susceptibility of surrounding medium.
The induced density of {\em sp}-band electrons is given as
\begin{eqnarray}
\label{back2}
\delta N_s({\bf r})= 
\frac{4\pi \delta n_s(r)}{(2l+1)r^{2l+1}},
 \end{eqnarray}

Integrating by parts, Eq.\ (\ref{poisson}) takes the form
\begin{eqnarray}
\label{poisson1}
\epsilon(r)\Psi({\bf r}) 
=\psi_0({\bf r}) + e^2\int d^3r' 
\frac{\delta N_s({\bf r'})}{|{\bf r}-{\bf r}'|}
\nonumber\\
+ \frac{\epsilon_d-1}{4\pi} \int d^3r' \nabla'\frac{1}{|{\bf
r}-{\bf r}'|}\cdot \nabla'\theta(R_d-r) \Psi({\bf r}')
\nonumber\\
+\frac{\epsilon_m-1}{4\pi} \int d^3r'
\nabla'\frac{1}{|{\bf r}-{\bf r}'|}\cdot \nabla'\theta(r-R) \Psi({\bf r}'),
\end{eqnarray}

where $\epsilon(r)= (\epsilon_d$, 1, $\epsilon _m$) for $r$ in the
intervals [$(0,R)$, $(R_d,R)$, $(R,\infty)$], respectively. We expand
$\Psi$ and $\delta N_s$ in terms of spherical harmonics and obtain,

\begin{eqnarray}
\label{poisson2}
\epsilon(r)\psi_{l}(r) =\psi_0({\bf r})  
\nonumber\\
+e^2 \int dr'r'^2 B_{l}(r,r')\delta N_s(r')
\nonumber\\
- \frac{\epsilon_d-1}{4\pi} R_d^2 \partial_{R_d} B_{l}(r,R_d)\psi_{l}(R_d)
\nonumber\\
+\frac{\epsilon_m-1}{4\pi} R^2 \partial_{R} B_{l}(r,R)\psi_{l}(R),
\end{eqnarray}

where 
\begin{equation}
\label{coulomb} 
B_{l}(r.r') =\frac{4\pi}{2l+1}\biggl[\frac{r'^l}{r^{l+1}} \,
\theta(r-r') +\frac{r^l}{r'^{l+1}} \, \theta(r'-r)\biggr]
\end{equation}

is the multipole term of the radial component of the Coulomb
potential. The above equation can be simplified to

\begin{eqnarray}
\label{poisson-nano}
%[1+(\epsilon_d-1)\theta(R_d-r)+(\epsilon_m-1)\theta(r-R)]
\epsilon(r) \psi_l(r) = \bar{\psi_l}(r)
%\nonumber\\
- \frac{\epsilon_d-1}{2l+1} \beta_l(r/R_d)\psi_l(R_d)
\nonumber\\
+ \frac{\epsilon_m-1}{2l+1} \beta_l(r/R) \psi_l(R),
\end{eqnarray}

where we introduced a shorthand notation
\begin{equation}\label{phi-s}
\bar{\psi_l}(r) = \psi_0({\bf r})+ e^2 \int dr'r'^2 B_l(r,r')\delta N_s(r'),
\end{equation}
and $\beta_l(r/R)=\frac{2l+1}{4\pi}R^2\partial_R B_l(R,r)$ is given by
\begin{equation}
\label{beta}
\beta_l(x) = lx^{-(l+1)}\, \theta (x-1) -(l+1)x^l \theta(1-x).
\end{equation}
The boundary values of $\psi$ can be obtained by matching $\psi(r)$
at $r= R_d, R$,
\begin{eqnarray}
\label{boundary}
(l\epsilon_d+l+1)\psi_l(R_d)+(l+1)a^l(\epsilon_m-1)\psi_l(R)
\nonumber\\
=(2l+1)\bar{\psi_l}(R_d)~~~~~~~~~~~\nonumber\\
l(\epsilon_d-1)a^{l+1}\psi_l(R_d)+(l\epsilon_m+\epsilon_m+1)\psi_l(R)
\nonumber\\
=(2l+1)\bar{\psi_l}(R)~~~~~~~~~~~
\end{eqnarray}
where $a=R_d/R$. Substituting $\psi_l(R_d)$ and $\psi_l(R)$ back into
Eq.\ (\ref{poisson-nano}), we arrive at
\begin{eqnarray}
\label{poisson-nano1}
\epsilon(r) \psi_l(r)= \bar{\psi_l}(r)
\nonumber\\
- \beta_l(r/R_d) \, \frac{\lambda_d}{\eta}\,
\Bigl[\bar{\psi_l}(R_d)-(l+1)a^l \lambda_m \bar{\psi_l}(R)\Bigr] 
\nonumber\\
+\beta_l(r/R) \, \frac{\lambda_m}{\eta}\,
\Bigl[\bar{\psi_l}(R)-la^{l+1}\lambda_d\bar{\psi_l}(R_d)\Bigr],
\end{eqnarray}

where the coefficients $\lambda$ are given by
\begin{eqnarray}
\label{lambda}
a=R_d/R, ~~~
\lambda_d=\frac{\epsilon_d-1}{l\epsilon_d+l+1},~~~
\lambda_m=\frac{\epsilon_m-1}{l\epsilon_m+\epsilon_m+1},
\nonumber\\
\eta=1-l(l+1)a^{2l+1}\lambda_d\lambda_m, ~~~~~~~~~~~~~
\end{eqnarray}

 Separating out $\delta n_s$-dependent
contribution, we arrive at TDLDA equation Ref.
\begin{eqnarray}
\label{tdlda2}
\delta n_s^{(l)}(r) = \int d r' r'^2 \Pi_s^{(l)} (r,r') 
\Bigl[w_0^{(l)}(r')+\delta w_0^{(l)}(r')\Bigr] 
\nonumber\\+
\int d r'  r'^2 \Pi_s^{(l)} (r,r') \Biggl[\int d r'' r''^2  
A(r',r'')\delta n_s^{(l)} (r'')
\nonumber\\ + V'_x(r')\delta n_s^{(l)} (r')\Biggr],~~~~~~~~~~~~~~~~~~~~~~~~~~
\end{eqnarray}
where  $\Pi_s ({\bf r}, {\bf
r}')$ is the polarization operator for noninteracting 
{\em sp}-electrons, $V'_x[n(r')]$ is the (functional) derivative of the
exchange-correlation potential and $n(r)$ is the ground-state electron
density. The latter is obtained in a standard way by solving Kohn-Sham
equations. Here we define  $w_0^{(l)}(r)=r^{l}/ \epsilon(r)$ and 

\begin{eqnarray}\label{delta-phi-0}
\delta w_0^{(l)}(r)&=&\frac{R^l}{\epsilon(r)} 
\Bigl[ -\beta_l(r/R_d) \lambda_d a^l(1-(l+1)\lambda_m)/\eta
\nonumber\\
 &+& \beta_l(r/R) \lambda_m(1-la^{2l+1}\lambda_d)/\eta \Bigr],
\nonumber\\
\delta w_s^{(l)}(r)&=&\int dr' r'^2 A(r,r')\delta n_s^{(l)}(r')
\end{eqnarray}
where kernel A(r,r') is a renormalized Coulomb potential $\Psi(r$) due to d-band and medium 
expressed as 
\begin{eqnarray}
\label{A}
A(r,r')= \frac{1}{\epsilon(r)} \Biggl[ B_{l}(r,r') -
\beta_l(r/R_d)\frac{\lambda_d}{\eta}\Biggl[B_{l}(R_d,r')
\nonumber\\
-a^{l}(l+1)\lambda_m B_{l}(R,r')\Biggr] + \beta_l(r/R)\frac{\lambda_m}{\eta}
\Biggl[B_{l}(R,r')
\nonumber\\
-la^{l+1}\lambda_d B_{l}(R_d,r') \Biggr]
 \Biggr].
\end{eqnarray}

The expression for quantum nanoparticle polarizability $\alpha^{q}=\alpha_d+\alpha_s$,
 can be obtained from the asymptotic for $r\gg R$
\begin{eqnarray}
\label{delta-phi-a}
\delta w_0^{(l)}(r) = \frac{4 \pi}{(2l+1)\epsilon_m r^{l+1}}\, 
\alpha_d,~~~~~~~~~~~
\nonumber\\
\delta w_s^{(l)}(r) = \frac{4 \pi}{(2l+1)\epsilon_m r^{l+1}}\,
 \alpha_s,~~~~~~~~~~
\end{eqnarray}

with
\begin{eqnarray}
\label{alpha2}
\alpha_d
=
\frac{l R^{2l+1}}{\eta}\Bigl[ \lambda_d a^{2l+1} (1-\lambda_m) - \lambda_m 
\Bigr],~~~~~~~~~~~~~~~~~~~~
\nonumber\\
\alpha_s
=\int_0^{\infty}dr'r'^{l+2}\delta n_s^{(l)}(r')~~~~~~~~~~~~~~~~~
\nonumber\\
-\frac{l(2l+1) R_d^{l+1} \lambda_d}{4\pi \eta} \int_0^{\infty}dr'r'^2\delta n_s^{(l)}(r')
\Biggl[B_{l}(R_d,r')
\nonumber\\
-a^{l}(l+1)
\lambda_m B_{l}(R,r')\Biggr]\nonumber\\ + 
\frac{l(2l+1) R^{l+1} \lambda_m}{4\pi \eta} \int_0^{\infty}dr'r'^2\delta n_s^{(l)}(r')
\Biggl[B_{l}(R,r')
\nonumber\\
-la^{l+1}\lambda_d B_{l}(R_d,r') \Biggr],~~~~~~~~~~~~~~
\end{eqnarray}
%

% BibTeX users please use
% \bibliographystyle{}
% \bibliography{}

\end{document}